\documentclass[aps,prl,twocolumn,superscriptaddress]{revtex4}
\usepackage{amsfonts}
\usepackage{amsmath}
\usepackage{amssymb}
\usepackage{graphicx}
\usepackage{color}
\usepackage{ulem}

\begin{document}

\title{Magnetic orders in a Fermi gas induced by cavity-field fluctuations}
\author{Jingtao Fan}
\affiliation{State Key Laboratory of Quantum Optics and Quantum Optics Devices, Institute
of Laser spectroscopy, Shanxi University, Taiyuan 030006, China}
\affiliation{Collaborative Innovation Center of Extreme Optics, Shanxi University,
Taiyuan, Shanxi 030006, China}
\author{Xiaofan Zhou}
\affiliation{State Key Laboratory of Quantum Optics and Quantum Optics Devices, Institute
of Laser spectroscopy, Shanxi University, Taiyuan 030006, China}
\affiliation{Collaborative Innovation Center of Extreme Optics, Shanxi University,
Taiyuan, Shanxi 030006, China}
\author{Wei Zheng}
\email{zhengwei8796@gmail.com}
\affiliation{T.C.M. Group, Cavendish Laboratory, J. J. Thomson Avenue, Cambridge CB3 0HE,
United Kingdom}
\author{Wei Yi}
\email{wyiz@ustc.edu.cn}
\affiliation{Key Laboratory of Quantum Information, University of Science and Technology
of China, CAS, Hefei, Anhui, 230026, China}
\affiliation{Synergetic Innovation Center of Quantum Information and Quantum Physics,
University of Science and Technology of China, Hefei, Anhui 230026, China}
\author{Gang Chen}
\email{chengang971@163.com}
\affiliation{State Key Laboratory of Quantum Optics and Quantum Optics Devices, Institute
of Laser spectroscopy, Shanxi University, Taiyuan 030006, China}
\affiliation{Collaborative Innovation Center of Extreme Optics, Shanxi University,
Taiyuan, Shanxi 030006, China}
\author{Suotang Jia}
\affiliation{State Key Laboratory of Quantum Optics and Quantum Optics Devices, Institute
of Laser spectroscopy, Shanxi University, Taiyuan 030006, China}
\affiliation{Collaborative Innovation Center of Extreme Optics, Shanxi University,
Taiyuan, Shanxi 030006, China}

\begin{abstract}
We study magnetic orders of fermions under cavity-assisted Raman couplings
in a one-dimensional lattice at half filling. The cavity-enhanced
atom-photon coupling introduces a dynamic long-range interaction between the
fermions, which competes with the short-range on-site interaction and leads
to a variety of magnetic orders. Adopting a numerical
density-matrix-renormalization-group method, we investigate the various
magnetic orders and map out the steady-state phase diagram. Interestingly,
as all the phase transitions take place outside the superradiant regime, the
magnetic orders are associated with cavity-field fluctuations with a
vanishing number of photons on the mean-field level.
\end{abstract}

\maketitle

Coherently driven atomic gases inside optical cavities have attracted much
research interest of late~\cite{Ritsch13}. In these systems, as the atoms
serve as a non-linear media between the external pumping and the cavity
fields, the cavity photons feed back on the atomic degrees of freedom,
effectively imposing a dynamic potential on the atoms. These dynamic
potentials are responsible for interesting non-equilibrium collective
dynamics and exotic steady states, which are the subjects of intensive
experimental and theoretical study~\cite%
{Baumann10,Mottl,Hemmerich,Landig16,Hruby,
Supersolid17,Goldstone17,Ritsch08,spinglass1,spinglass2,HabibianL13,dicketheory1,dicketheory2,dicketheory3,cavfermion1,cavfermion2,cavfermion3,Zheng17, Mekhov,Ritsch15,Dogra16,Zhai16,Zheng16,Corinna16,Ameneh16,Kyle17,Catalin17,Deng14,Bikash14,Dong15,Zhu16,AmenehT16,Pan15,Piazza17}%
.

Recently, a series of seminal experiments have demonstrated the impact of
different forms of dynamic potentials on atomic gases inside cavities~\cite%
{Landig16,Hemmerich,Mottl,Hruby,Baumann10,Supersolid17,Goldstone17}. A
prominent example is the observation of the supersolid phase transition in a
transversely pumped Bose-Einstein condensate coupled to a cavity~\cite%
{Baumann10,Supersolid17,Goldstone17}. In the experiment, as the cavity field
becomes superradiant, the back action of the photons induces a dynamic
superlattice potential, and drives the atoms into a self-organized steady
state. Further, it has been predicted theoretically~\cite%
{Mekhov,Dogra16,Ritsch15,Zhai16} and subsequently experimentally~\cite%
{Mottl,Landig16,Hemmerich,Hruby} verified that cavity-induced dynamic
long-range interactions can lead to a rich phase diagram for a Bose-Hubbard
model inside a cavity. Meanwhile, much theoretical effort has been dedicated
to the study of cavity-assisted dynamic gauge potentials, both abelian~\cite%
{Zheng16,Corinna16,Ameneh16,Kyle17,Catalin17} and non-abelian~\cite%
{Deng14,Bikash14,Dong15,Zhu16,AmenehT16,Pan15,Piazza17}, in atom-photon
ensembles, with the prospect of generating anomalous non-equilibrium
dynamics~\cite{Zheng16,Corinna16,Ameneh16}, or steady states with exotic
phases and correlations~\cite%
{Deng14,Bikash14,Catalin17,Kyle17,Pan15,AmenehT16,Piazza17}. Whereas most
previous studies focus on the superradiant regime, where the cavity field is
essentially in a coherent state, the cavity-field fluctuations should become
crucial away from superradiance. An intriguing question is then the
clarification of the impact of cavity-field fluctuations on atoms.

In this work, we show that fermions coupled to a cavity can develop
interesting magnetic phases under the dynamic long-range interaction driven
by cavity-field fluctuations. We focus on the steady state of a
two-component Fermi gas in a one-dimensional optical lattice coupled to a
cavity via the Raman transition. The fermions can be effectively described
by an extended Hubbard model with both the on-site and the dynamic
long-range interactions~\cite{Ritsch08}. As the long-range interaction
features spin-flipping processes, it competes with the spin-conserving
on-site interactions and leads to the emergence of magnetic correlations
and, consequently, various magnetic orders in the steady state.

Adopting the numerical density-matrix-renormalization-group (DMRG) method,
we map out the steady-state phase diagram and demonstrate that, as the
cavity parameters are tuned, the system is driven from an antiferromagnetic
state to a ferromagnetic state, with various anisotropic magnetic orders
lying in-between. As magnetic phase transitions typically take place in the
regime with a blue cavity detuning, where there is no superradiance, the
magnetic orders as well as the quantum phase transitions in-between are
driven by cavity-field fluctuations rather than the superradiance. Our work
is therefore in sharp contrast to previous theoretical and experimental
studies~\cite{Baumann10,Mottl,Hemmerich,Landig16,Hruby,
Supersolid17,Goldstone17,Ritsch08,spinglass1,spinglass2,HabibianL13,dicketheory1,dicketheory2,dicketheory3,cavfermion1,cavfermion2,cavfermion3,Zheng17, Mekhov,Ritsch15,Dogra16,Zhai16,Zheng16,Corinna16,Ameneh16,Kyle17,Catalin17,Deng14,Bikash14,Dong15,Zhu16,AmenehT16,Pan15,Piazza17}%
, where the focus has been on the phenomena induced by superradiance.

\begin{figure}[tbp]
\includegraphics[width=7cm]{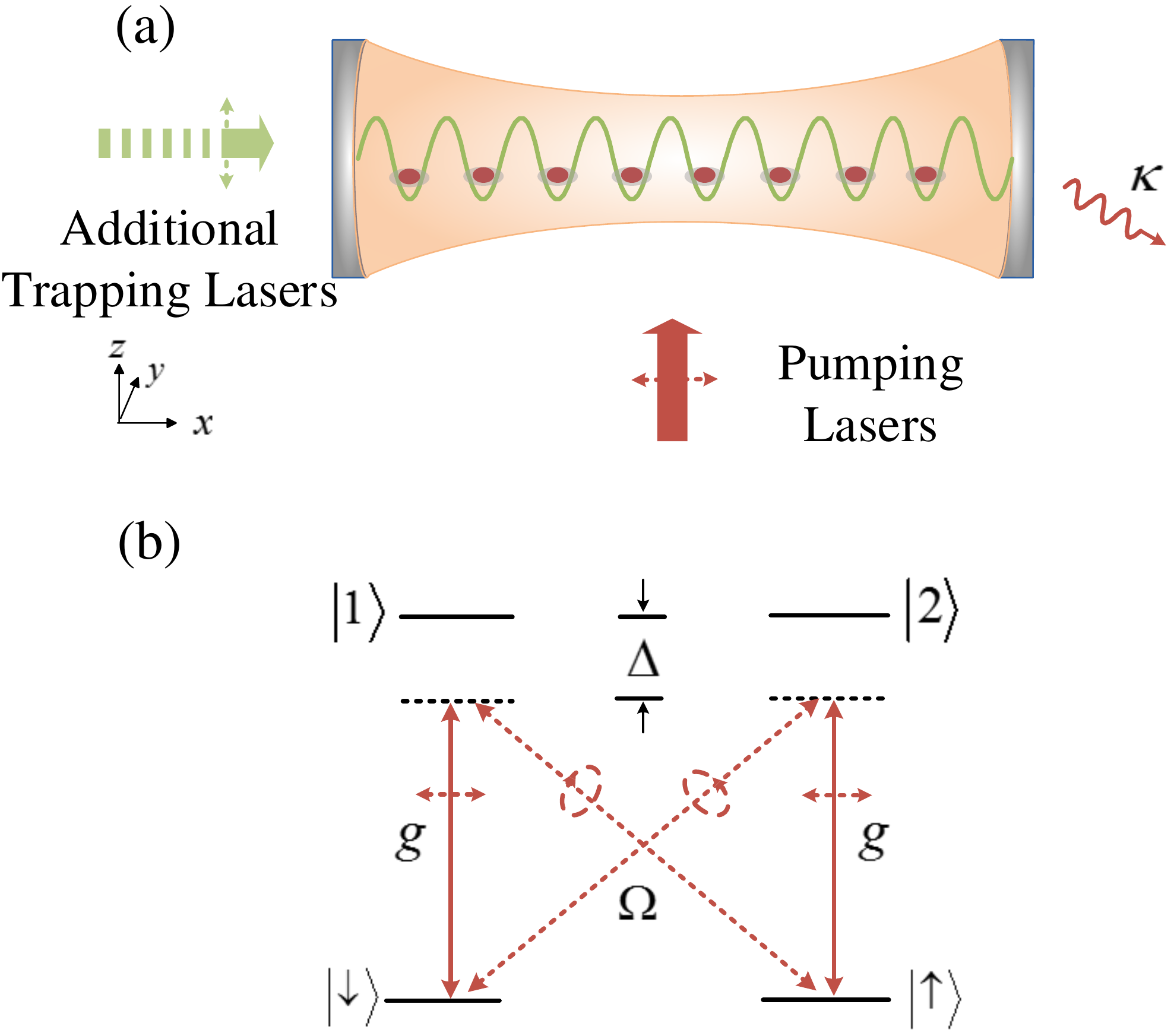}
\caption{(a) Fermions in a one-dimensional lattice potential are loaded into
a high-finesse optical cavity, which is subject to a linearly polarized
transverse pumping laser. (b) The pumping laser and the cavity field couple
two hyperfine states in two separate Raman processes. See main text for the
definition of different labels.}
\label{Level}
\end{figure}

\textit{Model:---} We consider a two-component Fermi gas in a
quasi-one-dimensional optical lattice potential inside a high-finesse
optical cavity, as shown in Fig.~\ref{Level}. While the lattice and the
cavity are aligned along the $x$-axis, the atoms are tightly confined in the
transverse directions so that only the atomic motion along the $x$-direction
is relevant. The cavity is subject to a transverse pumping laser, which,
together with the cavity field, couples the two hyperfine states ($%
\{\left\vert \downarrow \right\rangle ,\left\vert \uparrow \right\rangle \}$%
) of the fermions in two separate Raman processes (we take the $z$ direction
as the quantization axis). The cavity frequency $\omega _{c}$ is close to
that of the pumping laser ($\omega _{p}$), both of which are red-detuned
from the electronically excited states ($\{|1\rangle ,|2\rangle \}$) with
large single-photon detuning $\Delta \gg g,\Omega $. Here $g$ is the
single-photon Rabi frequency of the cavity field, and $\Omega $ is the Rabi
frequency of the pumping laser.

Adiabatically eliminating the excited states and adopting the tight-binding
approximation, the Hamiltonian of the system can be written as
\begin{align}
\hat{H}& =(\Delta _{c}-M_{1}\underset{j,\sigma }{\sum }\hat{n}_{j\sigma })%
\hat{a}^{\dag }\hat{a}-t\underset{j,\sigma }{\sum }\left( \hat{c}_{j\sigma
}^{\dag }\hat{c}_{j+1\sigma }+\text{H.c.}\right)  \notag \\
& +\eta \left( \hat{a}^{\dag }+\hat{a}\right) M_{0}\underset{j}{\sum }\left(
-1\right) ^{j+1}\left( \hat{c}_{j\uparrow }^{\dag }\hat{c}_{j\downarrow }+%
\hat{c}_{j\downarrow }^{\dag }\hat{c}_{j\uparrow }\right)  \notag \\
& +\frac{U_{s}}{2}\underset{j}{\sum }\hat{n}_{j\uparrow }\hat{n}%
_{j\downarrow },  \label{HCA}
\end{align}%
where $(\hat{a},\hat{a}^{\dag })$ are the field operators for the cavity
photons in the frame rotating with a frequency $\omega _{p}$, $c_{j\sigma
}^{\dag }$ ($c_{j\sigma }$) creates (annihilates) a fermion with spin $%
\sigma $ ($\sigma =\uparrow ,\downarrow $) on site $j$, and the density
operator $\hat{n}_{j\sigma }=\hat{c}_{j\sigma }^{\dag }\hat{c}_{j\sigma }$.
The cavity detuning is given by $\Delta _{c}=\omega _{c}-\omega _{p}$, and
the effective pumping strength $\eta =\nu g\Omega /\Delta $, with the
constant $\nu $ coming from transverse integrals~\cite{Pan15}. $U_{s}$ and $%
t $ are respectively the on-site interaction strength and the lattice
hopping rate. We also have $M_{1}=\frac{g^{2}}{\Delta }\int dxW_{j}^{\ast
}\cos ^{2}(k_{0}x)W_{j}$, and $M_{0}=\int dxW_{j}^{\ast }\cos (k_{0}x)W_{j}$%
, where $W_{j}$ is the Wannier function centered at site $j$ and $k_{0}$ is
the wave vector of the cavity field. In the Hamiltonian above, we have
neglected the Zeeman terms of the hyperfine spins, which corresponds to
vanishing two-photon detunings and a zero external magnetic field. Our main
results however should persist under a finite Zeeman field.

Taking the cavity decay into account, we may derive the Heisenberg equation
of motion for the cavity-field operators. As we seek the steady-state
solution, we further require $\partial _{t}\hat{a}=0$, which leads to
\begin{equation}
\hat{a}=\frac{\eta M_{0}}{i\kappa +\tilde{\delta}}\underset{j}{\sum }\left(
-1\right) ^{j+1}\left( \hat{c}_{j\uparrow }^{\dag }\hat{c}_{j\downarrow }+%
\hat{c}_{j\downarrow }^{\dag }\hat{c}_{j\uparrow }\right) ,  \label{AST}
\end{equation}%
where $\kappa $ is the cavity decay rate, the effective cavity detuning $%
\tilde{\delta}=NM_{1}-\Delta _{c}$, and $N$ is the total atom number.

An important observation is that the cavity field is associated with the
antiferromagnetic spin correlations along the $x$-direction in the steady
state, as $\hat{a}\propto \sum_{j}(-1)^{j+1}\hat{C}_{j}^{\dag }\sigma _{x}%
\hat{C}_{j}$, where $\sigma _{x}$ is the corresponding Pauli matrix and $%
C_{j}=(c_{j\uparrow },c_{j\downarrow })^{T}$. Whereas such a relation plays
a key role in driving the steady state of fermions into the
magnetically-ordered phases, this point becomes immediately clear if we
adiabatically eliminate the cavity field in the large-dissipation limit~\cite%
{Ritsch08}. The resulting effective Hamiltonian of the fermions is
essentially an extended Hubbard model with both on-site interactions and a
dynamic long-range interaction potential
\begin{align}
\hat{H}& =-t\underset{j,\sigma }{\sum }\left( \hat{c}_{j\sigma }^{\dag }\hat{%
c}_{j+1\sigma }+\text{H.c.}\right) +\frac{U_{s}}{2}\underset{j}{\sum }\hat{n}%
_{j\uparrow }\hat{n}_{j\downarrow }  \notag \\
& +\frac{U_{l}}{L}\left[ \underset{j}{\sum }\left( -1\right) ^{j+1}\left(
\hat{c}_{j\uparrow }^{\dag }\hat{c}_{j\downarrow }+\hat{c}_{j\downarrow
}^{\dag }\hat{c}_{j\uparrow }\right) \right] ^{2}.  \label{HTAT}
\end{align}%
Here $U_{l}/L=\left\vert \eta M_{0}\right\vert ^{2}\tilde{\delta}/(\tilde{%
\delta}^{2}+\kappa ^{2})$, where $L$ is the total number of lattice sites.
Note that while $t$ and $U_{s}$ can be tuned by adjusting the lattice
parameters, the long-range interaction strength $U_{l}$ can be tuned over a
wide range by adjusting parameters such as the pumping strength $\Omega $,
the effective cavity detuning $\tilde\delta $, and the atom-cavity coupling
rate $g$.

From the Hamiltonian (\ref{HTAT}), it is apparent that the competition of
the spin-preserving on-site interactions (characterized by $U_{s}$) and the
spin-flipping long-range interactions (characterized by $U_{l}$) can give
rise to interesting magnetically-ordered phases. This is particularly true
if we consider a repulsive on-site interaction with $U_{s}>0$, where it is
well known that an isotropic antiferromagnetic order is favored at half
filling when $U_{l}=0$~\cite{Scholl,Anderson,Lieb62,Lieb68,Lieb89}. In
contrast, for large and positive $U_{l}$, the formation of an
antiferromagnetic order along the $x$-direction would be hindered, and the
system should favor a ferromagnetic order along the $x$-direction.
Therefore, a quantum phase transition should occur between these limiting
cases. Further, as the dynamic long-range interaction breaks the SO(3)
symmetry of the original Hubbard model to an SO(2) rotational symmetry
around the $x$-axis, the magnetic orders should in general be anisotropic.

To clarify the impact of the dynamic long-range interaction on the magnetism
of the steady state, in the following, we perform numerical simulations
using the DMRG calculations, for which we retain 150 truncated states per
DMRG block and perform 20 sweeps with a maximum truncation error $\sim
10^{-5}$. We will focus on a half-filled lattice ($N/L=1$) with repulsive
on-site interactions $U_s>0$ under open boundary conditions.

\textit{Effect of the dynamic long-range interaction:---} The existence of
the magnetic order can be characterized by the static spin structure factor~%
\cite{Russell,Masao,Santos16,CCC}
\begin{equation}
S_{\alpha }(k)=\frac{1}{L}\underset{l,j}{\sum }e^{ik(l-j)}\left\langle \hat{s%
}_{l}^{\alpha }\hat{s}_{j}^{\alpha }\right\rangle ,
\end{equation}%
where $\hat{s}_{j}^{\alpha }=\frac{\hbar }{2}C_{j}^{\dag }\sigma _{\alpha
}C_{j}$, and $\sigma _{\alpha }$ ($\alpha =x,y,z$) are the Pauli matrices.
As the position of peaks in $S_{\alpha }(k)$ characterizes the spatial
variation of spin orientations projected into the $\alpha $-direction, peaks
at $k=\pm \pi $ and $k=0$ represent respectively antiferromagnetic and
ferromagnetic orders in the corresponding direction \cite%
{Russell,Masao,Santos16,CCC}. Alternatively, we can characterize the
magnetic order using the spin correlation function
\begin{equation}
C_{\alpha }(r)=\frac{1}{L}\underset{l}{\sum }\left\langle \hat{s}%
_{l}^{\alpha }\hat{s}_{l+r}^{\alpha }\right\rangle ,
\end{equation}%
where $r$ is the distance between different sites. The advantage of $%
C_{\alpha }(r)$ is that it offers an intuitive picture on the spatial
distribution of the spin correlations~\cite%
{Daul98,ZhouX,CCC,Boll16,Parsons16,Cheuk,Frederik17}. More specifically, for
an antiferromagnetic state, the sign of $C_{\alpha }(r)$ should alternate as
$r$ increases; while for phases with dominantly ferromagnetic correlations, $%
C_{\alpha }(r)$ should become purely positive~\cite%
{Daul98,Santos16,Boll16,Parsons16,Cheuk,Frederik17}.

\begin{figure}[tp]
\includegraphics[width=8.5cm]{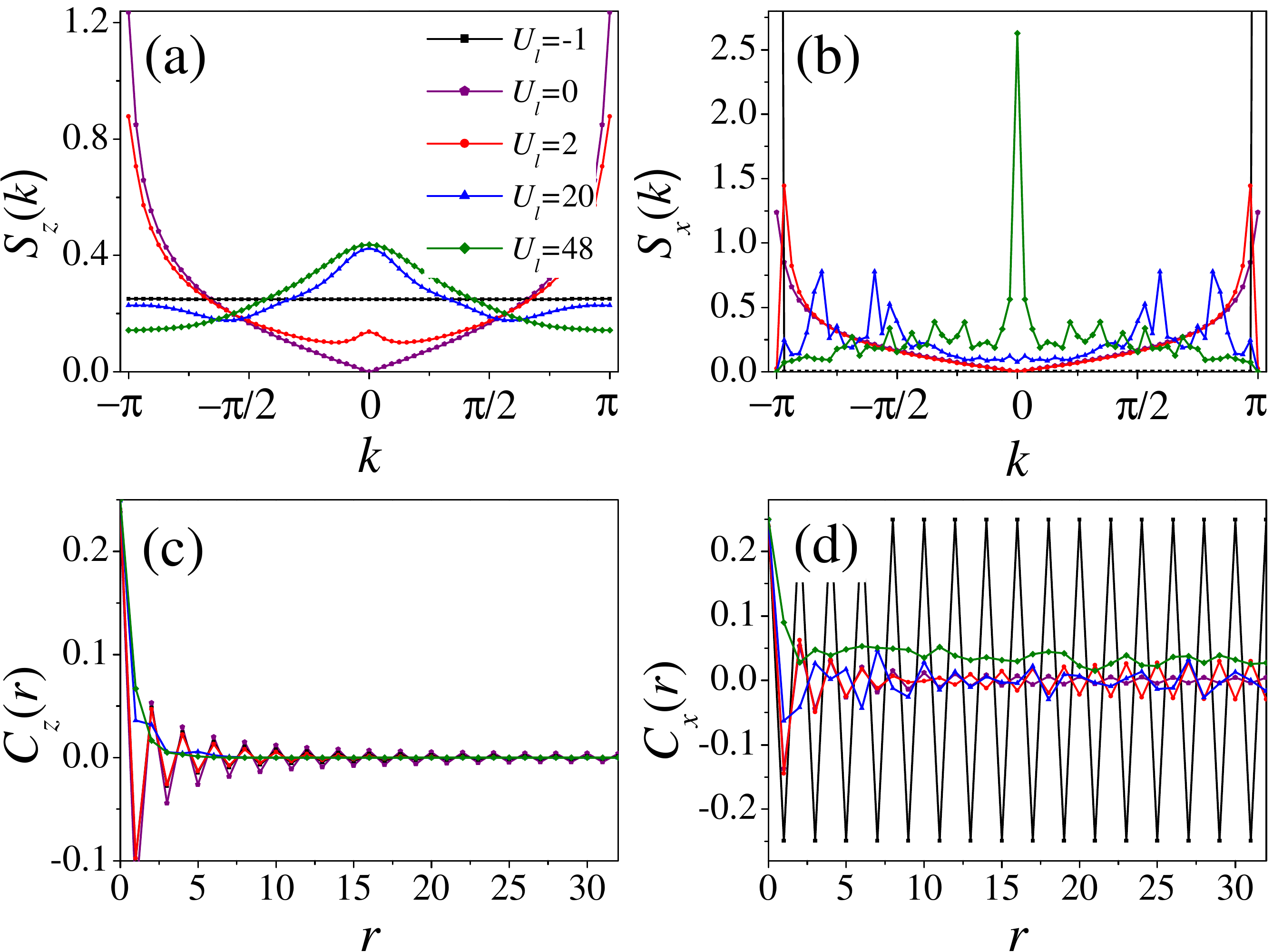} 
\caption{Spin structure factors (a) $S_{z}(k)$ and (b) $S_{x}(k)$, and spin
correlation functions (c) $C_{z}(r)$ and (d) $C_{x}(r)$ for systems with $%
t=0.1$, $L=64$, and a varying $U_{l}$.}
\label{Ul}
\end{figure}

We first study the variation of $S_{\alpha }(k)$ $\left( \alpha =x,z\right) $
with different values of $U_{l}$ at a fixed $t/U_{s}=0.1$. As illustrated in
Figs.~\ref{Ul}(a) and \ref{Ul}(b), when $U_{l}=0$, $S_{z}(k)$ and $S_{x}(k)$
peak identically at $k=\pm \pi $, which is consistent with the existence of
an isotropic antiferromagnetic order in the ground-state of a repulsive
Hubbard model at half filling. This is further confirmed by the spin
correlations, as $C_{z}(r)$ and $C_{x}(r)$ oscillate identically around zero
as $r$ increases [see Figs.~\ref{Ul}(c) and \ref{Ul}(d)]. For finite $U_{l}$
however, the system is only isotropic in the transverse directions ($y$-$z$
plane), as the long-range interaction breaks the SO(3) symmetry. This is
reflected in the drastically different peak structures along the $x$- and
the $z$-directions when $U_{l}\neq 0$.

For $U_{l}<0$, the system features anisotropic antiferromagnetic orders
along the $x$- and the $z$-directions, as $S_{z}(k)$ and $S_{x}(k)$ still
peak at $k=\pi $ but with different peak structures. The spin correlations
also behave differently along the two directions. The oscillations in $%
C_{x}(r)$ appear to be undamped, suggesting long-range antiferromagnetic
order. This is in contrast to the power-law decay of the $C_{z}(r)$
oscillations, which indicates quasi-long-range order in the $z$-direction.

As $U_{l}$ becomes positive, a new peak at $k=0$ immediately emerges in $%
S_{z}(k)$, which suggests the building up of ferromagnetic correlations. The
peak at $k=0$ eventually becomes higher than that at $k=\pi $, as $U_{l}$ is
increased beyond a critical value $U_{l}/U_{s}\approx 16$, which we
associate with a transition from antiferromagnetism to ferromagnetism in the
transverse direction. Such a transition can be confirmed by the spin
correlations, as $C_z(r)$ becomes purely positive beyond the critical $U_l$.

The situation along the $x$-direction is more complicated. As soon as $U_{l}$
becomes positive, the peak in $S_{x}(k)$ is shifted away from $k=\pi $. This
suggests that the spin orientations become non-collinear in the $x$%
-direction, which we identify as an incommensurate antiferromagnetic order~%
\cite{LXP17}. In the incommensurate antiferromagnetic state, $C_x(r)$ still
features oscillations around zero, but with periods incommensurate with that
of the lattice. At larger $U_{l}$, the competition between the on-site and
the long-range interactions gives rise to multiple peaks in $S_{x}(k)$,
which eventually merge into a single one at $k=0$ when $U_{l}$ is increased
above a critical value $U_{l}/U_{s}\approx 28$. Therefore, when the
long-range interaction is strong enough, the system becomes ferromagnetic
along both the $x$- and the transverse directions, where both $C_z(r)$ and $%
C_x(r)$ become purely positive.

\begin{table}[t]
\caption{The correspondence between $\protect\vartheta _{\protect\alpha }$
and different magnetic phases, where $\protect\vartheta _{\protect\alpha }$
denotes the position of the highest peak in $S_{\protect\alpha }(k)$ ($%
\protect\alpha =x,z$).}
\label{table1}%
\begin{tabular}{ccccc}
\hline\hline
Position of peak & AF & FM & AF$_z$-IAF$_x$ & FM$_z$-IAF$_x$ \\ \hline
$\vartheta _{z}$ & $\pm \pi $ & 0 & $\pm \pi $ & 0 \\
$\,\vartheta _{x}$ & $\pm \pi $ & 0 & $\neq $0,$\pm \pi $ & $\neq $0,$\pm
\pi $ \\ \hline\hline
\end{tabular}%
\newline
\end{table}

\begin{figure}[tp]
\includegraphics[width=8cm]{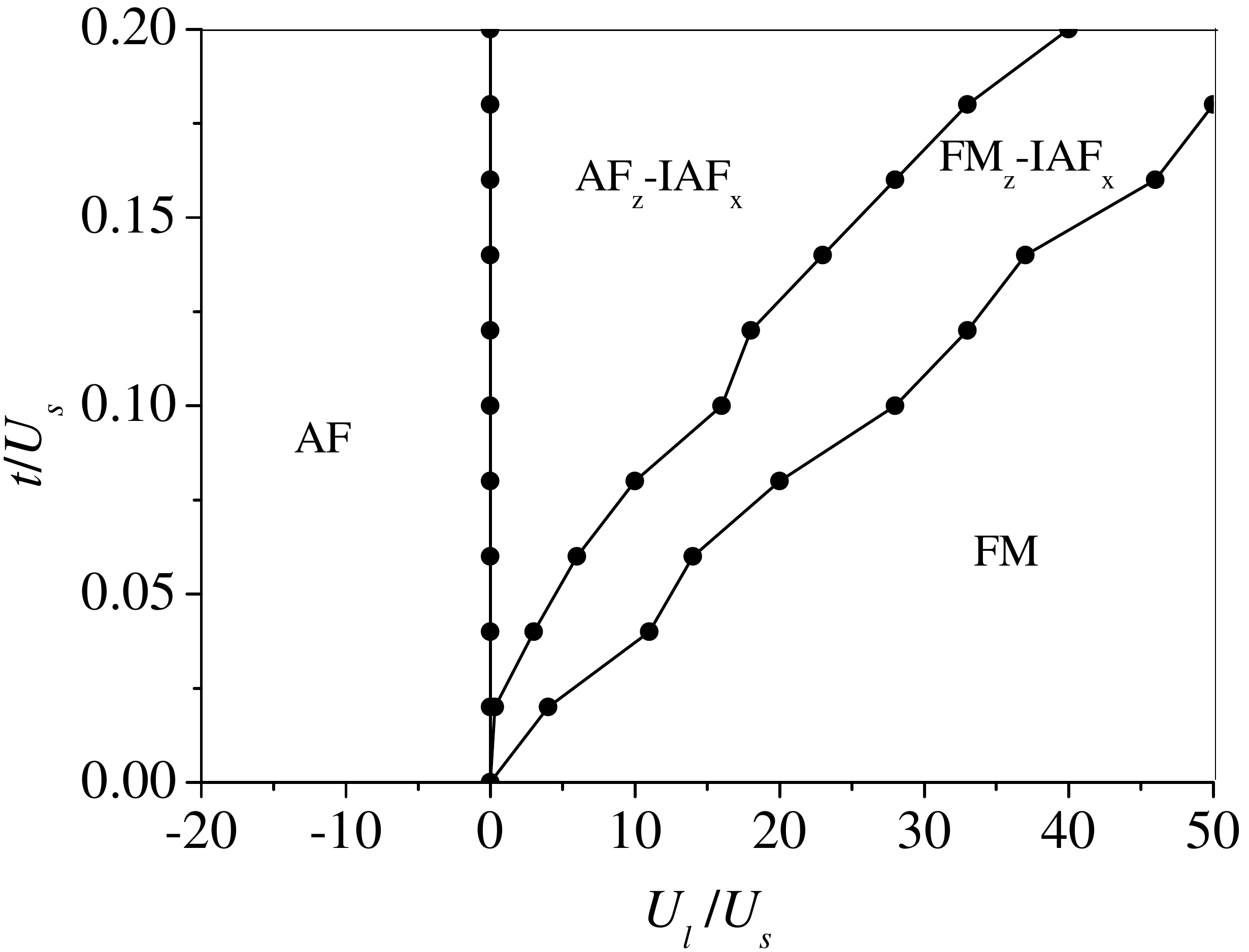}
\caption{The phase diagram in the $U_{l}-t$ plane for a system with $L=64$.
AF, FM, and IAF correspond to antiferromagnetic state, ferromagnetic state,
and incommensurate antiferromagnetic state, respectively, with the
subscripts indicating the direction of the magnetic order. The definitions
of the different phases are listed in Table \protect\ref{table1}. }
\label{phaseDiagram}
\end{figure}

With the understanding above, we map out the phase diagram of the system in
the $U_{l}-t$ plane. The magnetic orders are identified from the peak
locations in the corresponding structure factors (see Table \ref{table1} for
detailed descriptions). For example, the phase boundaries for the
ferromagnetic orders are determined by requiring the peak at $k=0$ being
equal in height with the highest peak elsewhere. As shown in Fig.~\ref%
{phaseDiagram}, in the transverse directions, the system changes from an
antiferromagnetic phase to a ferromagnetic phase at positive $U_{l}$. While
in the $x$-direction, the steady-state is antiferromagnetic for $U_{l}<0$,
incommensurate antiferromagnetic for intermediate $U_{l}$, and ferromagnetic
in the large $U_{l}$ limit. The magnetic order is indeed anisotropic in
general. Note that as the phase diagram is obtained for a finite-size
lattice, we have numerically confirmed its qualitative validity in the
thermodynamic limit $L\rightarrow \infty $ using a finite-size-scaling
analysis~\cite{supp}.

\begin{figure}[tbp]
\includegraphics[width=7cm]{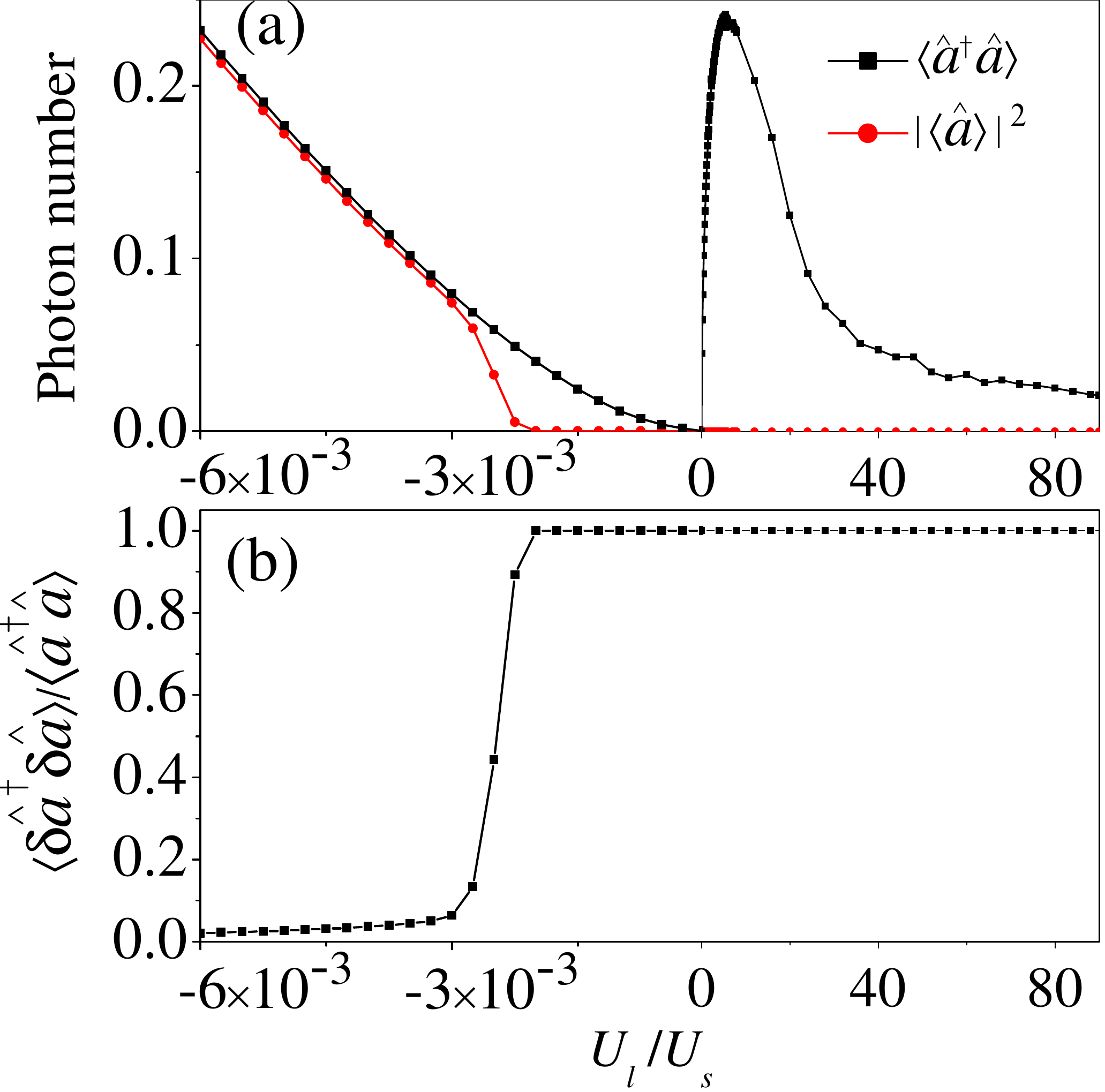} 
\caption{(a) The cavity photon number and (b) the cavity-field fluctuation $%
\left\langle \protect\delta \hat{a}^{\dag }\protect\delta \hat{a}%
\right\rangle /\left\langle \hat{a}^{\dag }\hat{a}\right\rangle $ as a
function of $U_{l}$, where we have taken $t/U_{s}=0.1$, $\tilde{\protect%
\delta}/U_{s}=1$, and $L=64$.}
\label{photonUl}
\end{figure}

\textit{Cavity-field fluctuations:---} As the cavity field is associated
with the antiferromagnetic correlations according to Eq.~(\ref{AST}), it
serves as the driving force behind magnetic transitions. To further clarify
the role and the behavior of cavity photons throughout the phase
transitions, in Fig.~\ref{photonUl}, we plot the number of cavity photons $%
\langle \hat{a}^{\dag }\hat{a}\rangle $ in the steady state with varying $%
U_{l}$. For comparison, we also show the square of the mean cavity field $%
|\langle \hat{a}\rangle |^{2}$, which should be close to $\langle \hat{a}%
^{\dag }\hat{a}\rangle $ when the mean-field approximation is valid and the
cavity field can be approximated by a coherent state.

In Fig.~\ref{photonUl}(a), at a first glance, we can identify a superradiant
transition, where $|\langle \hat{a}\rangle |^{2}$ becomes finite, for an
effectively red-detuned cavity $(\tilde{\delta}<0, U_{l}<0)$. As $U_{l}$
decreases further, $|\langle \hat{a}\rangle |^{2}$ increases and rapidly
approaches $\langle \hat{a}^{\dag }\hat{a}\rangle $. This indicates that the
cavity field can be described by a coherent state when the system is in the
superradiant regime.

In contrast, in the region where the cavity is effectively blue-detuned $(%
\tilde{\delta}>0,U_{l}>0)$ and where all the magnetic phase transitions take
place, the cavity is not superradiant, as $|\langle \hat{a}\rangle |^{2}$
remains vanishingly small. The cavity is therefore dominated by fluctuations
as $|\langle \hat{a}\rangle |^{2}$ deviates considerably from the photon
number $\langle \hat{a}^{\dag }\hat{a}\rangle $. In Fig.~\ref{photonUl}(b),
we characterize cavity-field fluctuations with $\langle \delta \hat{a}^{\dag
}\delta \hat{a}\rangle /\langle \hat{a}^{\dag }\hat{a}\rangle $ $(\delta
\hat{a}=\hat{a}-\langle \hat{a}\rangle $), which vanishes in a coherent
state and approaches unity in the case of large cavity fluctuations. Our
results are consistent with a recent experiment~\cite{Hemmerich2015}, where
the absence of superradiance has been reported for a blue-detuned cavity.
Therefore, both the incommensurate antiferromagnetic and the ferromagnetic
orders are induced by the cavity-field fluctuations instead of superradiance.

With increasing $U_{l}$, the cavity-photon number undergoes a non-monotonic
change, with a peak situated at $U_{l}/U_{s}\approx 5.5$. To understand this
behavior, we can relate the cavity photon number to the structure factor
\begin{align}
\left\langle \hat{a}^{\dag }\hat{a}\right\rangle =4\frac{U_{l}}{\tilde{\delta%
}}S_{x}(k=\pi ).  \label{PHO}
\end{align}%
According to Eq.~(\ref{PHO}), the steady-state photon number is collectively
determined by both $U_{l}$ and the static structure factor $S_{x}(k=\pi )$.
Therefore, an anti-parallel spin configuration would feed back positively on
the cavity photon number and vice versa. In the absence of the atom-photon
coupling ($U_{l}=0$), the photon number is equal to zero, and the fermions
are in the antiferromagnetic state. A small and positive $U_{l}$ is not
sufficient to break the intrinsic antiferromagnetic order of the system,
which in turn gives rise to an increase of the photon number. By increasing $%
U_{l}$ further, the ferromagnetic configuration starts to dominate, which
would then decrease the photon number. In the large $U_{l}$ limit, the
average photon number monotonically approaches zero, as the steady state
acquires an anisotropic ferromagnetic order. We note that while the lack of
superradiance for $U_l>0$ can be confirmed by a mean-field calculation with $%
U_s=0$~\cite{supp}, the characterization of the steady-state in this region
is clearly beyond the mean-field approach.

\textit{Discussions:---} We have shown that magnetic phases and phase
transitions can be induced in a one-dimensional Fermi gas by cavity-field
fluctuations away from the superradiant regime. Such a behavior is
drastically different from previous studies focusing on the effects of
superradiance, where the mean-field approach is still applicable. The
magnetic phase transitions lead to signals in the spin dynamical structure
factor, which can be detected by measuring the photons leaking out of the
cavity~\cite{Esslinger2015}. Alternatively, the magnetic orders can be
probed by constructing the spin correlation function from spin-resolved
\textit{in situ} measurements~\cite{Parsons16,Greiner2017}. It will be
interesting to consider situations in higher dimensions, in which a richer
phase diagram is expected.

\textit{Acknowledgement:---} This work is supported partly by the National
Key R\&D Program of China under Grants No.~2017YFA0304203 and
No.~2016YFA0301700; the NKBRP under Grant No.~2013CB922000; the NSFC under
Grants No.~60921091, No.~11374283, No.~11434007, No.~11522545, and
No.~11674200; \textquotedblleft Strategic Priority Research
Program(B)\textquotedblright\ of the Chinese Academy of Sciences under Grant
No.~XDB01030200; the PCSIRT under Grant No.~IRT13076; SFSSSP; OYTPSP; and
1331KYC. Jingtao Fan and Xiaofan Zhou contributed equally to this work.

\newpage
\begin{widetext}
\appendix

\renewcommand{\thesection}{\Alph{section}}
\renewcommand{\thefigure}{S\arabic{figure}}
\renewcommand{\thetable}{S\Roman{table}}
\setcounter{figure}{0}
\renewcommand{\theequation}{S\arabic{equation}}
\setcounter{equation}{0}

\section{Supplemental Materials}

In this Supplemental Materials, we demonstrate the
finite-size scaling of the DMRG results, we also characterize the
superradiant phase transition under the mean-field approximation. The notation here follows those in the main text.

\section{Finite-size scaling}

We show the finite-size scaling of four representative points in the phase diagram
in Fig.~\ref{scaling}. It is apparent that the critical points
remain finite in the thermodynamic limit $L\longrightarrow \infty $,
which confirms the validity of the phase diagram in the main text.

\begin{figure}[h]
\hskip 1.5cm\includegraphics[width=10cm]{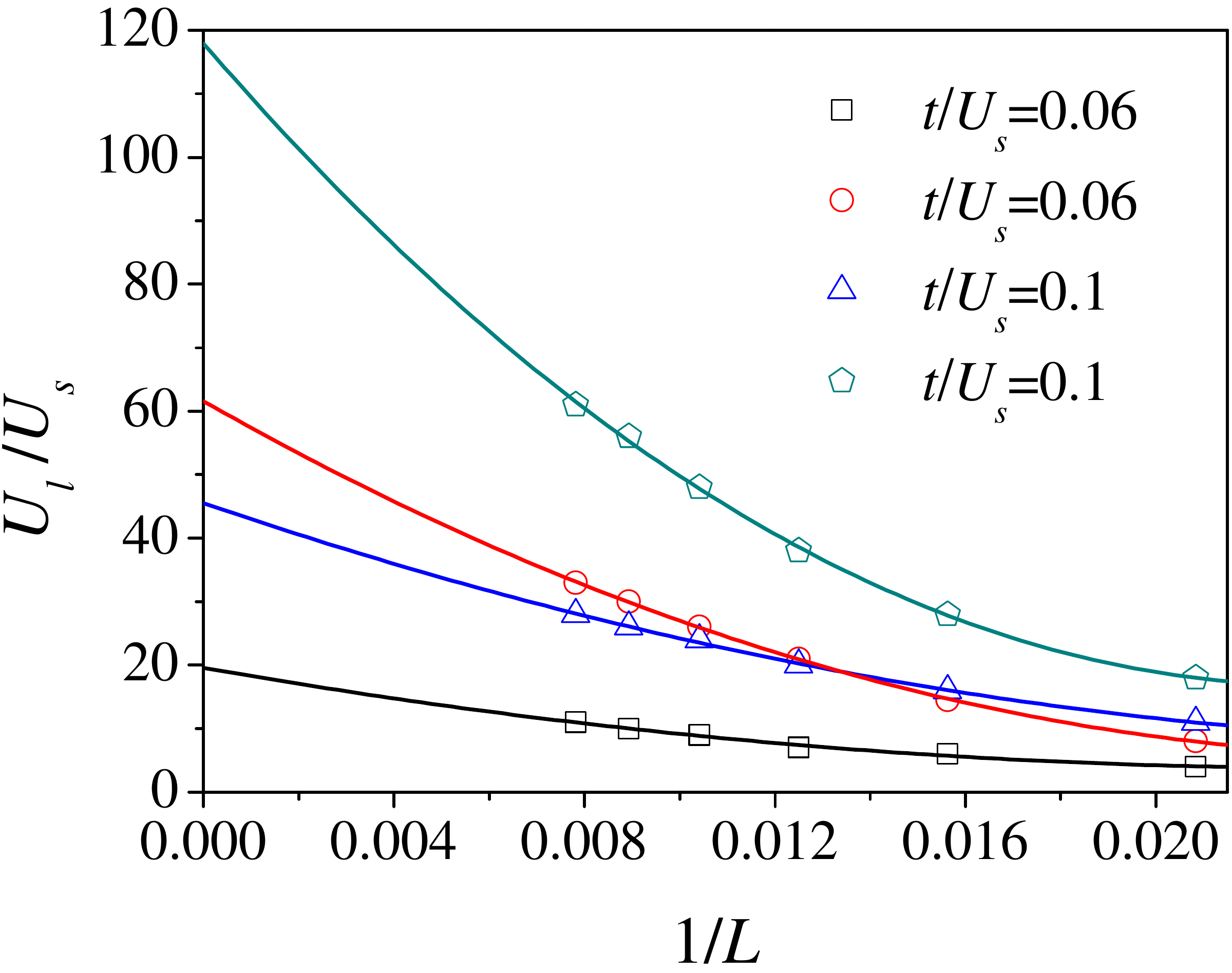}\newline
\caption{The finite-size scaling of the critical interaction strength $U_{l}$
calculated for two different values of $t/U_{s}$. The scaling function is
a second-order polynomial in $1/L$.}
\label{scaling}
\end{figure}

\section{Self-consistent mean-field calculation of cavity field}

In the absence of the on-site interaction ($U_s=0$), the superradiance of the
cavity field can be characterized under the mean-field approximation.

We start from the tight-binding Hamiltonian Eq.~(1) in the main text. Employing a local unitary transformation $\hat{c}_{i\uparrow }\rightarrow
(-1)^{i+1}\hat{c}_{i\uparrow }$, the Hamiltonian becomes%
\begin{equation}
\hat{H}=-\delta \hat{a}^{\dag }\hat{a}-t\underset{j,\sigma }{\sum }\left( \hat{c}_{j\sigma }^{\dag }\hat{c}_{j+1,\sigma }+\text{H.c.}\right)
+\eta \left( \hat{a}^{\dag }+\hat{a}\right) M_{0}\underset{j}{\sum }\left(
\hat{c}_{j\uparrow }^{\dag }\hat{c}_{j\downarrow }+\hat{c}_{j\downarrow
}^{\dag }\hat{c}_{j\uparrow }\right) .  \label{HIF}
\end{equation}%
The Heisenberg equation for $\hat{a}$ is
\begin{equation}
\partial _{t}\hat{a}=(i\tilde{\delta}-\kappa )\hat{a}-i\eta M_{0}\underset{j}%
{\sum }\left( \hat{c}_{j\uparrow }^{\dag }\hat{c}_{j\downarrow }+\hat{c}%
_{j\downarrow }^{\dag }\hat{c}_{j\uparrow }\right),
\end{equation}%
where the parameters $\tilde{\delta} $ is defined in the main text. Under the mean-field approximation $
\left\langle \hat{a}\right\rangle =\alpha $ and using the steady-state condition $\partial_t\alpha=0$, we have
\begin{equation}
\alpha =\frac{\eta M_{0}}{i\kappa +\tilde{\delta}}\underset{j}{\sum }%
\left\langle \hat{c}_{j\uparrow }^{\dag }\hat{c}_{j\downarrow }+\hat{c}%
_{j\downarrow }^{\dag }\hat{c}_{j\uparrow }\right\rangle .  \label{Alpha}
\end{equation}%
Note that the cavity is assumed to be in a coherent state under the mean-field approximation, with the average photon number given by $|\alpha|^2$.

The cavity field $\alpha$ can be calculated self-consistently as the following:
(i) diagonalize the Hamiltonian (\ref{HIF}) from an initial value of the cavity-field $\alpha_0$
; (ii) determine the chemical potential from the number equation $N=\sum_{j,\sigma }\left\langle \hat{c}%
_{j\sigma }^{\dag }\hat{c}_{j\sigma }\right\rangle $; (iii) update the
cavity field $\alpha $ with Eq.~(\ref{Alpha}); (iv) replace $\alpha_0$ with the current value of $\alpha$ and repeat
steps (i-iii) until $\alpha$ converges.

In Fig.~\ref{PhotonS}, we show the calculated average photon number $|\alpha|^2$ (blue). For comparison,
we have also plotted $|\langle \hat{a}\rangle|^2$ (red) and the photon number
$\langle \hat{a}^{\dag}\hat{a}\rangle$ (black) from the DMRG calculations. From the mean-field results, it is apparent that
the system is superradiant for $U_l/t<-0.3$. In the superradiant regime, the mean-field photon number $|\alpha|^2$ agrees well with the photon number $|\langle \hat{a}\rangle|^2$ from the DMRG calculations. However, for $U_l>0$, the system is not superradiant as both $|\alpha|^2$ and $|\langle \hat{a}\rangle|^2$ vanish. The finite photon number $\langle \hat{a}^{\dag}\hat{a}\rangle$ is therefore the result of cavity-field fluctuations, whose effects on the fermions are beyond the mean-field description. Importantly, the overall picture here is consistent with Fig.~4 in the main text, where a finite on-site interaction is considered.

\begin{figure}[tp]
\centering
\hskip 1cm\includegraphics[width=10cm]{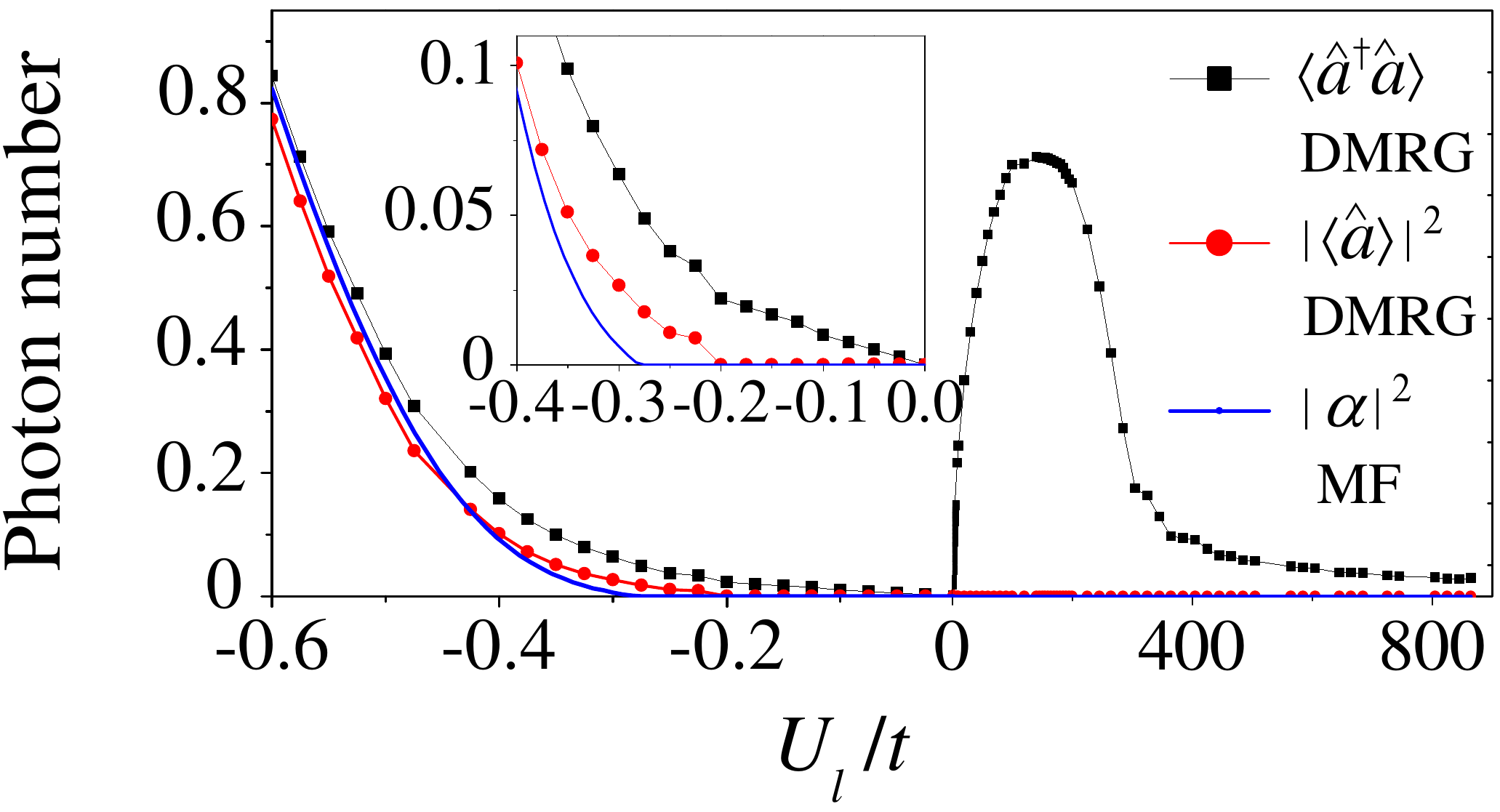} \newline
\caption{Comparison of the cavity photon number $|\alpha|^2$ from the mean-field (MF) calculation (blue), as well as the square of the cavity field $|\langle \hat{a}\rangle|^2$ (red) and the photon number $\langle \hat{a}^{\dag}\hat{a}\rangle$ (black) from the DMRG calculations. We focus on the case with $\tilde{\delta}/t=10$, $U_{s}=0$, and $L=64$.}
\label{PhotonS}
\end{figure}

\end{widetext}

\end{document}